\RequirePackage{fix-cm}
\documentclass[pdftex,twocolumn,epjc3]{svjour3}
\smartqed  
\RequirePackage{graphicx}
\RequirePackage[colorlinks,citecolor=blue,urlcolor=blue,linkcolor=blue]{hyperref} 
\usepackage{color}
\usepackage{multirow}
\usepackage{amssymb}     
\usepackage{marvosym}    
\usepackage{upgreek}     
\usepackage{amsmath}
\newcommand{\kgyr}        {{kg$\cdot$yr}}

\newcommand{\mum}         {{$\upmu$m}}
\newcommand{\mus}         {{$\upmu$s}}

\newcommand{\qbb}         {{$Q_{\beta\beta}$}}

\newcommand{\thalftwo}    {${T^{2\nu}_{1/2}}$}
\newcommand{\thalfmajo}   {${T^{0\nu \chi }_{1/2}}$}
\newcommand{\nmez}        {${\cal M}^{0\nu}$}
\newcommand{\nmet}        {${\cal M}^{2\nu}$}

\newcommand{\onbb}        {{$0\nu\beta\beta$}}
\newcommand{\onbbchi}     {{$0\nu\beta\beta\chi$}}
\newcommand{\nnbb}        {{$2\nu\beta\beta$}}
\newcommand{\twonu}       {{$2\nu\beta\beta$}}

\newcommand{\etal}        {\textit{et al.}}

\newcommand{\gerda}       {\textsc{Gerda}}
\newcommand{\G}           {{\mbox{\textsc{Gerda}}}}
  
\newcommand{\IGEX}        {{\mbox{\textsc{Igex}}}}

\newcommand{\HdM}         {\mbox{\textsc{HdM}}}  
\newcommand{\HDM}         {\mbox{\textsc{HdM}}}

\newcommand{\KAMLAND}     {{\mbox{\textsc{KAMland}}}}

\newcommand{\NEMO}        {{\mbox{\textsc{Nemo}}}}

\newcommand{\EXO}         {{\mbox{\textsc{Exo}}}}

\newcommand{\geant}       {\textsc{Geant4}}
\newcommand{\GEANT}       {\textsc{\mbox{{Geant}}}}

\newcommand{\mage}        {\textsc{MaGe}}
\newcommand{\gelatio}     {\textsc{Gelatio}}

\newcommand{\gess}        {{$^{76}$Ge}}

\newcommand{\exposure}    {\mbox{$\cal E$}}

\journalname{Eur. Phys. J. C}

\begin{document}

\title{
Results on $\beta\beta$ decay with emission of two neutrinos or 
Majorons in $^{76}$Ge from \mbox{\protect{\textsc{GERDA}}} Phase~I}

\author{
   M.~Agostini\thanksref{TUM} \and
   M.~Allardt\thanksref{DD} \and
   A.M.~Bakalyarov\thanksref{KU} \and
   M.~Balata\thanksref{ALNGS} \and
   I.~Barabanov\thanksref{INR} \and
   N.~Barros\thanksref{DD} \and
   L.~Baudis\thanksref{UZH} \and
   C.~Bauer\thanksref{HD} \and
   N.~Becerici-Schmidt\thanksref{MPIP} \and
   E.~Bellotti\thanksref{MIBF,MIBINFN}
% ...
}

\authorrunning{the \textsc{Gerda} collaboration}
\thankstext{alsoMIPT}{\emph{also at:} Moscow Inst. of Physics and Technology,
  Russia} 
\thankstext{alsoIUN}{\emph{also at:} Int. Univ. for Nature, Society and
    Man ``Dubna'', Dubna, Russia} 
\thankstext{corrauthor}{\emph{Correspondence},
                                email: gerda-eb@mpi-hd.mpg.de}
\institute{%
INFN Laboratori Nazionali del Gran Sasso and Gran Sasso Science Institute,
               Assergi, Italy\label{ALNGS} \and
INFN Laboratori Nazionali del Sud, Catania, Italy\label{CAT} \and
Institute of Physics, Jagiellonian University, Cracow, Poland\label{CR} \and
Institut f{\"u}r Kern- und Teilchenphysik, Technische Universit{\"a}t Dresden,
      Dresden, Germany\label{DD} \and
Joint Institute for Nuclear Research, Dubna, Russia\label{JINR} \and
Institute for Reference Materials and Measurements, Geel,
     Belgium\label{GEEL} \and
Max Planck Institut f{\"u}r Kernphysik, Heidelberg, Germany\label{HD} \and
Dipartimento di Fisica, Universit{\`a} Milano Bicocca,
     Milano, Italy\label{MIBF} \and
INFN Milano Bicocca, Milano, Italy\label{MIBINFN} \and
Dipartimento di Fisica, Universit{\`a} degli Studi di Milano e INFN Milano,
    Milano, Italy\label{MILUINFN} \and
Institute for Nuclear Research of the Russian Academy of Sciences,
    Moscow, Russia\label{INR} \and
Institute for Theoretical and Experimental Physics,
    Moscow, Russia\label{ITEP} \and
National Research Centre ``Kurchatov Institute'', Moscow, Russia\label{KU} \and
Max-Planck-Institut f{\"ur} Physik, M{\"u}nchen, Germany\label{MPIP} \and
Physik Department and Excellence Cluster Universe,
    Technische  Universit{\"a}t M{\"u}nchen, M{\"u}nchen, Germany\label{TUM}
                     \and
Dipartimento di Fisica e Astronomia dell{`}Universit{\`a} di Padova,
    Padova, Italy\label{PDUNI} \and
INFN  Padova, Padova, Italy\label{PDINFN} \and
Physikalisches Institut, Eberhard Karls Universit{\"a}t T{\"u}bingen,
    T{\"u}bingen, Germany\label{TU} \and
Physik Institut der Universit{\"a}t Z{\"u}rich, Z{\"u}rich,
    Switzerland\label{UZH}
}

\date{Received: date / Accepted: date}

\maketitle

\abstract{A search for neutrinoless $\beta\beta$ decay processes accompanied
  with Majoron emission has been performed using data collected during Phase~I
  of the GERmanium Detector Array (\G) experiment at the Laboratori Nazionali
  del Gran Sasso of INFN (Italy).  Processes with spectral indices $n$ = 1, 2,
  3, 7 were searched for. No signals were found and lower limits of the order
  of 10$^{23}$~yr on their half-lives were derived, yielding substantially
  improved results compared to previous experiments with $^{76}$Ge.  A new
  result for the half-life of the neutrino-accompanied $\beta\beta$ decay of
  $^{76}$Ge with significantly reduced uncertainties is also given, resulting
  in $T^{2\nu}_{1/2} = (1.926 \pm 0.095)\cdot10^{21}$~yr.
}

\section{Introduction}
\label{sec:intro}
Neutrinoless double beta (\onbb) decay is regarded as the gold-plated process
for probing the fundamental character of neutrinos.  Observation of this
process would imply total lepton number violation by two units and that
neutrinos have a Majorana mass component.  Although the main focus of
experimental efforts lies on the detection of \onbb\ decay mediated by light
Majorana neutrino exchange, there are also many other proposed mechanisms
which are being searched for.  Some exotic models predict \onbb\ decays
proceeding through the emission of a massless Goldstone boson, called
Majoron. Predictions of different models depend on its transformation
properties under weak isospin, singlet~\cite{chi80}, doublet~\cite{san88} and
triplet~\cite{gel81}.  Precise measurements of the invisible width of the Z
boson at LEP~\cite{LEP2006} greatly disfavour triplet and pure doublet models.
Several new Majoron models have been developed subsequently in which the
Majoron carries leptonic charge and cannot be a Goldstone
boson~\cite{burg93,burg94} or in which the \onbb\ decay proceeds through the
emission of two Majorons~\cite{bam95}.

All these models predict different shapes of the two emitted electrons' summed
energy spectrum. The predicted spectral shapes are essentially defined by the
phase space of the emitted particles:
\begin{equation}
 \frac {dN} {dK} \sim G \sim (Q_{\beta\beta} - K)^{n}
\label{spectral_n}
\end{equation}    
where $K$ is the summed energy of the two electrons, $G$ is the phase space,
$Q_{\beta\beta}$ is the $Q$ value of the \onbb\ decay and $n$ is the spectral
index of the model.  Single Majoron emitting $\beta\beta$ decays can be
roughly divided into three classes, $n=1$, $n=2$, and $n=3$.  Double Majoron
emitting decays can have either $n=3$ or $n=7$.  Their characteristic spectral
shapes differ from that of two-neutrino $\beta\beta$ decay (\nnbb), for
which $n=5$. This allows for discrimination between the processes.

Experimental searches for $\beta\beta$ decay mediated by emission of one or
two Majorons (\onbbchi) have been performed by the
Heidelberg-Moscow experiment (\HdM) for $^{76}$Ge~\cite{HdM,HdM-latest}; by
\NEMO-2 and \NEMO-3 for $^{100}$Mo, $^{116}$Cd, $^{82}$Se, $^{96}$Zr,
$^{130}$Te ~\cite{NEMO2006,NEMO2000}; by ELEGANT~V for
$^{100}$Mo~\cite{ELEGANT}; by DAMA~\cite{DAMA} and 
 by \KAMLAND-Zen
for $^{136}$Xe~\cite{KamLAND-Zen}.  None of these experiments have seen an
excess of events that could be interpreted as a Majoron signal; they reported
lower limits on the half-lives of the processes that involve Majoron emission.

The \nnbb\ decay process conserves lepton number and is independent of the
nature of the neutrino. It has been detected for eleven nuclides so far, with
measured half-lives ($T^{2\nu}_{1/2}$) in the range of $7 \times 10^{18} - 2
\times 10^{24}$\,yr~\cite{Barabash2010,Tretyak1995,Tretyak2002}.  The
knowledge of $T^{2\nu}_{1/2}$ allows for extraction of the nuclear matrix
element, \nmet, which can provide some constraints on that of \onbb\ decay,
\nmez, if the evaluations of ${\cal M}$ for the two processes are performed
within the same model~\cite{Rodin2006,Simkovic2008}.

This paper reports on the search for neutrinoless double beta decay of
$^{76}$Ge with Majoron emission (\onbbchi) and a new analysis of the half-life
of the \nnbb\ decay of $^{76}$Ge using data collected by the \G~experiment
during its Phase~I.  \nnbb\ decay is a well established and previously
observed process, while \onbbchi\ decay is a hypothetical one.  In the first
case the half-life is extracted, while for the second one a limit is set. This
leads to slightly different approaches in the analyses leading to different
data sets and background components being used.

\section{The 
\mbox{\protect{\textsc{GERDA}}} experiment}
\label{sec:gerda}
The main aim of the \G~experiment~\cite{gerda_tec:2013} at the Laboratori
Nazionali del Gran Sasso (LNGS) of INFN in Italy is to search for \onbb\ decay
of $^{76}$Ge.  The core of the setup is an array of high-purity germanium
(HPGe) detectors made from isotopically modified material with $^{76}$Ge
enriched to $\sim$86\,\% ($^{enr}$Ge), mounted in low-mass copper supports
(holders) and immersed in a 64~m$^3$ cryostat filled with liquid argon (LAr).
The LAr serves as cooling medium and shield against external backgrounds.  The
shielding is complemented by water in a tank of 10~m in diameter which is
instrumented with photomultipliers to detect Cherenkov light generated in
muon-induced showers~\cite{gerda_tec:2013}.

The array of HPGe detectors is arranged in strings.  Each string is enclosed
with a cylinder, made from 60~\mum\ thick Cu foil, called mini-shroud, to
mitigate the background coming from the decay of $^{42}$Ar present in the
LAr. Moreover, in order to prevent contamination from radon within the
cryostat, a cylinder, made from 30~\mum\ thick Cu foil, called radon-shroud,
separates the central part of the cryostat, where the detectors are located,
from the rest.  The HPGe detector signals are read out with custom-made charge
sensitive preamplifiers optimized for low radioactivity, which are operated
close to the detectors in the LAr. The analog signals are digitized with 100
MHz Flash ADCs (FADC) and analyzed offline. If one of the detectors has an
energy deposition above the trigger threshold (40-100~keV), all channels are
read out.  Reprocessed $p$-type coaxial detectors from the \HDM~\cite{HdM1997}
and \IGEX~\cite{IGEX1999} experiments were operated together with Broad Energy
Germanium (BEGe) type detectors manufactured by
Canberra~\cite{canberra,begeprod}.

\begin{figure}[b]
\begin{center}
\includegraphics[width=.57\columnwidth]{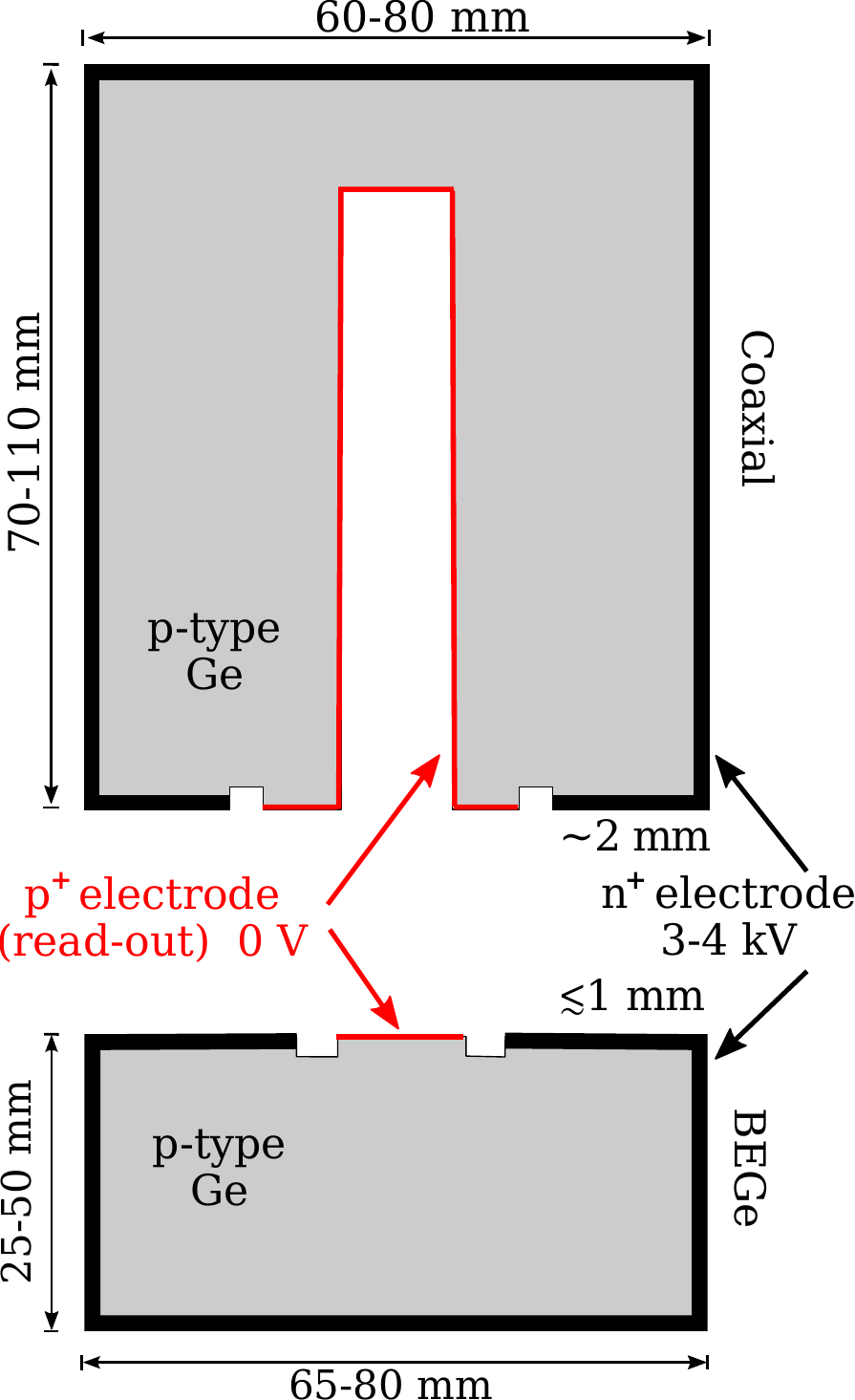}
\caption{\label{fig:coaxBegeReadout}
         Schematic sketch of a coaxial HPGe detector (top) and a BEGe detector
         (bottom) with their different surfaces and dead layers (drawings not
         to scale), adapted from Ref.~\cite{background-paper}.
}
\end{center}
\end{figure}

As explained in section~\ref{sec:background}, some background components have
different effects on the two detector types due to their peculiar geometry.  A
schematic drawing of a coaxial detector type is shown in the top part of
Fig.~\ref{fig:coaxBegeReadout}, while the lower part depicts that for a BEGe
type detector.

\section{Data taking and data selection}
\label{sec:data}
Phase~I data taking lasted from November 9, 2011, to May 21, 2013.  The total
exposure collected comprises 19.2~\kgyr\ for the coaxial detectors and
2.4~\kgyr\ for the BEGe detectors.  In this paper, the entire exposure
collected by the BEGe detectors (BEGe data set) and 17.9~\kgyr\ from the
coaxial detectors (golden data set) are
used~\cite{background-paper,gerda:0nubb}. For the coaxial detectors, a data
set collected for 1.3~\kgyr\ exposure during a restricted time period around
the deployment of the BEGe detectors is discarded due to a higher background
level.  Also one of the coaxial detectors, RG2, is not considered for the data
analysis starting from March 2013, as its high voltage had to be reduced below
depletion voltage due to increased leakage current.  The energy calibration of
the detectors was performed using the information from dedicated calibration
runs. For these calibration runs, three $^{228}$Th sources were lowered to the
vicinity of the detectors.  The stability of the energy scale was monitored by
performing such calibration runs every one or two weeks. Moreover, the
stability of the system was continuously monitored by injecting charge pulses
into the test input of the preamplifiers.  Using physics data, the
interpolated FWHM values at \qbb\ averaged with the exposure are (4.8 $\pm$
0.2)~keV for the coaxial detectors and (3.2 $\pm$ 0.2)~keV for the BEGe
detectors.

All steps of the offline processing of the \gerda\ data were performed within
the software framework \gelatio~\cite{gelatio}.  The energy deposited in each
detector was extracted from the respective charge pulse by applying a
approximate Gaussian filter~\cite{gaussfilter}.  Non-physical events, such as
discharges, cross-talk and pick-up noise events, were rejected by quality cuts
based on the time position of the rising edge, the information from the
Gaussian filter, the rise time and the charge pulse height, which must not
exceed the dynamic range of the FADCs.  Pile-up and accidental coincidences
were removed from the data set using cuts based on the baseline slope, the
number of triggers and the position of the rising edge. The rate of pile-up
and accidental coincidence events is negligible in the \gerda\ data due to the
extremely low event rate.  The loss due to mis-classification by the quality
cuts was $<$0.1\,\% for events with energies above 1~MeV.  All events that
come within 8~\mus\ of a signal from the muon veto were rejected.  Finally, only
events that survive the detector anti-coincidence cut were considered. This
means, that all events with an energy deposition $>50$~keV in more than one
detector in the array were not taken into account.  Since \nnbb\ and
\onbbchi\ events release their energy within a small volume inside the
detectors, almost no signal events were lost by this cut, while a part of the
$\gamma$-induced background events were rejected.

\section{Analysis Strategy}
\label{sec:strategy}
The two analyses described in this paper are different in the sense that for
\nnbb\ decay a parameter is extracted for a well established and known
process, while in the case of the search for \onbbchi\ decay limits for a
hypothetical process are set.  In order to minimize the systematic
uncertainties for the extraction of the \thalftwo\ it is favorable to use a
well defined and controlled subset of the data and to use only well identified
background processes. For \onbbchi\ limit setting it is favorable to maximize
the exposure and to take into account all known possible background processes
that can not be unambiguously detected but could mimic \onbbchi\ decay.

For the \thalftwo\ analysis the golden data set (17.9~\kgyr) with the coaxial
detectors is used in order to have a large data sample obtained in well
controlled experimental conditions.  The Majoron analysis uses both the golden
data set and the BEGe data set for a total exposure of 20.3~\kgyr\ in order to
maximize the sensitivity.

The background model for the \thalftwo\ analysis uses a minimal number of
components, assuming all sources near to the
detectors~\cite{background-paper,becerici}.  For the Majoron analysis, an
expanded model is used~\cite{hemmer}, taking into account also additional
medium and far distant positions for some of the sources.  This becomes
necessary when searching for rare processes such as Majoron emission, where
all possible sources of background which could simulate the exotic process
have to be considered. Therefore, even the slight differences resulting, for
example from a variation of the source position, have to be evaluated.

In both analyses, the experimental spectra of the coaxial and BEGe detectors
are analyzed using the Bayesian Analysis Toolkit (\textsc{Bat})~\cite{bat2}.

\section{The background model}
\label{sec:background}
The background sources considered in the models were identified by their
prominent structures in the energy spectra and were also expected on the basis
of material screening measurements.  The spectral shapes of individual
background contributions were obtained by using a detailed implementation of
the experimental setup in the Monte Carlo (MC) simulation framework
\mage~\cite{mage}.  A Bayesian spectral fit of the measured energy spectrum
with the simulated spectra was performed in an energy range from 570~keV up to
the end of the dynamic range at 7500~keV.  The low energy limit is motivated
by the $\beta$-decay of $^{39}$Ar, which gives a large contribution up to its
$Q_{\beta}$-value of 565~keV.

The following background components were used for the extraction of the
$T^{2\nu}_{1/2}$ (minimum model in Refs.~\cite{background-paper,becerici}):
(1) $^{76}$Ge \twonu\ decay, (2) $^{214}$Bi, $^{228}$Ac, $^{228}$Th, $^{60}$Co
and $^{40}$K decays in the close vicinity of the detectors ($<$2~cm,
represented by decays in the detector holders in the MC simulation), (3)
decays of $^{60}$Co inside the detectors, constrained by the maximum expected
activity from their cosmogenic activation history, (4) $^{42}$K decays in LAr
assuming a uniform distribution, (5) $\alpha$-model that accounts for $\alpha$
decays originating from $^{210}$Po and $^{226}$Ra contaminations on the
$p^{+}$ surface of the detectors as well as from $^{222}$Rn in the LAr, and
finally (6) $^{214}$Bi decays on the $p^{+}$ surface, constrained by the
estimated $^{226}$Ra activity from the $\alpha$-model.

The parameters of all components besides the constrained ones were given a
flat prior probability distribution.  There are no strong correlations between
the model parameters since all considered background components have
characteristic features such as $\gamma$-ray lines or peak-like structures at
different energies.  The ratios of the $\gamma$-ray line intensities from the
individual considered background sources suggest contaminations dominantly in
locations close to the detectors. Hence, the minimum model takes into account
only the close-by source locations.  Nevertheless, the screening measurements
indicate contaminations of materials in farther locations as well.  An
additional contribution can come from $^{42}$K decays at or near the detector
$n^+$ surfaces (see Fig.~\ref{fig:coaxBegeReadout}) with a specific activity
higher than that for the uniform distribution assumption.  This component is
the dominating one for the BEGe data set, as the thinner dead layer thickness
of BEGes of roughly 1~mm allows penetration of the electrons emitted in the
decay of $^{42}$K to the active volume, while for coaxial detectors the dead
layer thickness of $\sim$2~mm efficiently shields this background component.

The spectral shapes of the contributions from the background sources without
significant multiple $\gamma$ peaks at different source locations differ only
marginally. This makes it impossible to pinpoint the exact source locations
given the available statistics of the measured spectra. Therefore, variations
of the source locations for the considered decays were taken into account when
evaluating the systematic uncertainty on \thalftwo.

For the Majoron analysis additional background components were
used~\cite{hemmer}, including also medium and far distant contributions.  For
the coaxial detectors $^{42}$K on the $n^{+}$ and on the $p^{+}$ contacts was
added to the list of the close sources of the previous background model.  For
medium distances, i.e. between 2~cm and 50~cm from the detectors,
contributions from the following sources were added: $^{214}$Bi, $^{228}$Th
and $^{228}$Ac.  A $^{228}$Th contamination was chosen as a representative for
far distant sources (above 50~cm).  Whenever possible, screening measurements
were used to constrain the lower limit of the expected background events.

In the Majoron analysis, also the data collected with the BEGe diodes were
used in order to maximize the exposure.  Consequently, the background model
developed for these detectors was used~\cite{background-paper,hemmer}.  The
same close, medium and far distant sources as for the coaxial detectors were
used.  $^{68}$Ge was added as internal source.  This was necessary in order to
take into account the cosmic activation of the germanium due to the recent
production of these diodes.

\section{Determination of the half-life of \twonu\ decay}
\label{sec:twonu}
\subsection{Analysis}
\label{subsec:twonuAnalysis}
The \thalftwo\ of \twonu\ decay of \gess\ was determined considering the
golden data set of Phase~I, amounting to an exposure of 17.9~\kgyr, and using
the background model prediction for the contribution of the \twonu\ spectrum
to the overall energy spectrum.  Details of the background analysis can be
found in Ref.~\cite{becerici}.
  
The global fit for the background modeling was performed on the summed energy
spectrum of the coaxial detectors using a bin width of 30~keV.  Thus, the
scaling parameter of the \twonu\ spectrum in the model, $N^{\rm
  {fit}}_{2\nu}$, gives the number of events in the \twonu\ spectrum in the
fit window of 570--7500~keV for all detectors.  Using this result for the
number of measured \twonu\ events, the half-life is calculated as
\vspace{5pt}
\begin{eqnarray}\nonumber
  \label{eq:twonu:ThalfDerive}
T^{2\nu}_{1/2} & = & \frac { \left( \rm{ln}\,2 \right) \,
                             N_{A} }{ m_{\rm{enr}}\,\,N^{\rm {fit}}_{2\nu} } 
\sum_{i =1}^{N_{det}} M_{i}\,t_{i}\,f_{76, i} \left[ f_{AV, i}\,\,
                            \varepsilon^{\rm {fit}}_{AV,i} \right. \\
 && \left.  \qquad \qquad \qquad \qquad +\, (1-f_{AV, i})\,\,
                                \varepsilon^{\rm {fit}}_{DL,i} \right],
\end{eqnarray} 
\vspace*{5pt}

\noindent where $N_{A}$ is Avogadro's constant and $m_{\rm{enr}} = 75.6$~g is
the molar mass of the enriched material.  The summation runs over all the
detectors ($N_{det}$) considered in the data set.  All detector related
parameters like the detector mass ($M_i$), the time of the data taking for
each detector ($t_i$), the fraction of \gess\ atoms ($f_{76,i}$), the active
volume fraction ($f_{AV,i}$), and the detection efficiencies in the active
volume ($\varepsilon^{\rm{fit}}_{AV,i}$) and in the dead layer
($\varepsilon^{\rm{fit}}_{DL,i}$) are taken into account separately for the
individual detectors.  All values are listed in Table~\ref{tab:detValues}.
The efficiency $\varepsilon^{\rm{fit}}_{AV,i}$
($\varepsilon^{\rm{fit}}_{DL,i}$) corresponds to the probability that a
\twonu\ decay taking place in the active volume (dead layer) of the detector
deposits detectable energy in the fit window considered for the background
model.  The detection efficiencies, on average $\varepsilon^{\rm{fit}}_{AV} =
0.667$ and $\varepsilon^{\rm{fit}}_{DL} = 0.011$, are obtained through
dedicated MC simulations. The statistical uncertainty due to the number of
simulated events is on the order of 0.1\,\%.

\begin{table}[t] \footnotesize
  \centering
  \caption{  \label{tab:detValues}
        Parameters for the coaxial detectors (upper part) and for the BEGe
        detectors (lower part): live time, $t$, total mass, $M$, the fraction
        of \gess\ atoms, $f_{76}$, and the active volume fraction, $f_{AV}$.
        For the coaxial detectors, the first uncertainty on $f_{act}$ is the
        uncorrelated part, the second one the correlated contribution.  The
        values for $M$, $f_{76}$ and $f_{AV}$ are taken from
        Ref.~\cite{background-paper}.
}
  \vspace{5pt}
  \begin{tabular}{l c c c c}
    \hline
    detectors & $t$  & $M$ & $f_{76}$ & $f_{AV}$   \\
              & [days] & [kg]  & [$\%$]  & [$\%$]     \\
    \hline
    \multicolumn{5}{c}{enriched coaxial detectors} \\   
    \hline
    ANG2 & 485.5 & 2.833 & $86.6\pm 2.5$ & $87.1\pm 4.3\pm 2.8 $ \\
    ANG3 & 485.5 & 2.391 & $88.3\pm 2.6$ & $86.6\pm 4.9\pm 2.8 $ \\
    ANG4 & 485.5 & 2.372 & $86.3\pm 1.3$ & $90.1\pm 4.9\pm 2.9 $ \\ 
    ANG5 & 485.5 & 2.746 & $85.6\pm 1.3$ & $83.1\pm 4.0\pm 2.7 $ \\
    RG1  & 485.5 & 2.110 & $85.5\pm 1.5$ & $90.4\pm 5.2\pm 2.9 $ \\ 
    RG2  & 384.8 & 2.166 & $85.5\pm 1.5$ & $83.1\pm 4.6\pm 2.7 $ \\ 
    \hline
    \multicolumn{5}{c}{enriched BEGe detectors} \\   
    \hline
    GD32B & 280.0 & 0.717 & $87.7\pm 1.3$ & $89.0 \pm 2.7$ \\ 
    GD32C & 304.6 & 0.743 & $87.7\pm 1.3$ & $91.1 \pm 3.0$ \\ 
    GD32D & 282.7 & 0.723 & $87.7\pm 1.3$ & $92.3 \pm 2.6$ \\ 
    GD35B & 301.2 & 0.812 & $87.7\pm 1.3$ & $91.4 \pm 2.9$ \\ 
    \hline                                          
  \end{tabular}
\end{table}

The background model resulted in a scaling parameter of $N^{\rm {fit}}_{2\nu}
= 25690~^{+310}_{-330}$ for the \twonu\ spectrum, which is the best fit
parameter. The uncertainty is given by the smallest 68\,\% probability
interval of the marginalized posterior probability distribution. Using this
result, the half-life derived according to Eq.~\ref{eq:twonu:ThalfDerive} is
\begin{equation}
\label{eq:twonu:analysis}
T^{2\nu}_{1/2} = (1.926\,^{ +0.025}_{ -0.022}) \cdot 10^{21}~\rm{yr}~.
\end{equation}

\subsection{Systematic Uncertainties}
\label{subsec:twonuSyst}

The systematic uncertainties affecting the results for $T^{2\nu}_{1/2}$ were
grouped into the three categories {\it (i) detector parameters and fit model},
{\it (ii) MC simulation}, and {\it (iii) data acquisition and selection}.  The
contributions to the total systematic uncertainty on $T^{2\nu}_{1/2}$ are
summarized in Table~\ref{tab:systUncertainty}.

\begin{table*}
  \centering
   \caption{ \label{tab:systUncertainty}
                 Contributions to the systematic uncertainty on
                 \thalftwo\ taken into account in this work.  The total
                 systematic uncertainty is obtained by combining the
                 individual contributions in quadrature.
}
  \vspace{5pt}
  \begin{tabular*}{0.65\textwidth}{@{\extracolsep{\fill}} l c c @{}}
    \hline
    Item & \multicolumn{2}{c}{Uncertainty on \thalftwo\ } \\
         & \multicolumn{2}{c}{ [\%]} \\
\hline
  Active \gess\ exposure			& $\pm 4$	& \\
  Background model components			& $^{+ 1.4}_{- 1.2}$& \\ 
  Binning                                           & $\pm 0.5$ & \\
  Shape of the \twonu\ spectrum			& $< 0.1$	& \\
\hline
  Subtotal fit model				& 		& $\pm4.3$ \\
\hline
  Precision of the Monte Carlo geometry model	& $\pm 1$	& \\
  Accuracy of the Monte Carlo tracking		& $\pm 2$	& \\
\hline
  Subtotal Monte Carlo simulation		& 		& $\pm 2.2$ \\
\hline
  Data acquisition and handling			& 		& $< 0.1$ \\
\hline
  Total						& 		& $\pm 4.8$ \\  
\hline 
   \end{tabular*}
\end{table*}

\noindent{\it (i) detector parameters and fit model}

\begin{itemize}
\item 
The systematic uncertainty on the active \gess\ exposure ($\exposure_{AV,
  76}$) was determined using a MC approach. $\exposure_{AV, 76}$ is defined as
\begin{equation}
 \label{eq:cht:twonu:systematic:AVandEnrFrac:actExposure}
\exposure_{AV, 76} = \sum_{i=1}^{N_{det}} M_{i} t_{i} f_{AV,i} f_{76,i} ~.
\end{equation} 
For evaluating its uncertainty, the parameters of the individual detectors
were randomly sampled from Gaussian distributions with mean values and
standard deviations according to the corresponding values listed in Table
\ref{tab:detValues}.  The correlated terms for $f_{AV}$ were also taken into
account.  The uncertainty on the live time $t$ is 0.3\,\%, whereas the total
detector masses are known with good accuracy (uncertainty smaller than
0.1\,\%).  The calculation yields $\exposure_{AV, 76} = (13.45 \pm
0.54)$~kg$\cdot$yr.  The uncertainty of 4\,\% is driven by the uncertainties
on $f_{AV}$ and $f_{76}$, which mainly affect the number of \gess\ nuclei in
the active volume of the detectors, with a relatively smaller impact on the
detection efficiency for the background sources.

\item 
The reference background model used for determining \thalftwo\ accounts only
for the dominant source locations in the setup.  The systematic uncertainty
due to the choice of the background model components was evaluated by
repeating the global fit with alternative models, which account for different
source locations for all the background sources considered in the reference
model. The model that accounts for $^{228}$Th and $^{228}$Ac contributions
also in the radon-shroud instead of only in the holders results in a 1.4\,\%
longer \thalftwo. The same increase occurs if $^{40}$K in the radon-shroud is
added to the model components. The model including the contribution from
$^{214}$Bi in the radon-shroud in addition to the $p^{+}$ surface and holders
yields a 0.7\,\% longer \thalftwo.  In all the cases mentioned above, the
contribution from background in the \twonu\ spectrum region increases, since
the peak-to-Compton ratio of the $\gamma$-rays decreases for farther source
locations leading to longer \thalftwo\ estimates.  Excluding contributions
from very close source locations, like $^{214}$Bi on the $p^{+}$ surface and
$^{60}$Co on the germanium, results in a smaller increase of the best
\thalftwo\ estimate.  In this case, the contributions from these components
are compensated by $^{214}$Bi and $^{60}$Co decays in the holders,
respectively. Consequently, the source locations are moved further out with
respect to the reference model.  Consistently, the models that include
additional contributions from close source locations yield a decrease in the
\thalftwo\ value, e.g. including $^{214}$Bi in LAr close to the $p^{+}$
surface (-1.0\,\%) or $^{42}$K on the $n^{+}$ (-1.2\,\%) and $p^{+}$
(-0.6\,\%) surfaces.  Comparing alternative background models to the reference
one, the deviations in the \thalftwo\ result range between -1.2\,\% and
+1.4\,\%.
 
\item 
For the standard fit, a bin width of 30~keV was used for the data and MC
energy spectra.  In order to take into account the systematic uncertainty
related to binning effects, the fit was repeated twice using bin widths of 10
and 50~keV.  The bin width of 10~keV was chosen in order to minimize as much
as possible the bin size taking into account the energy resolution of
$\approx$4.5~keV of the coaxial detectors and the necessity to have enough
statistics in all bins.  Above 50~keV, peak structures are washed out, leading
to a deterioration of the fit.  The deviations in the \thalftwo\ result range
between -0.5\,\% and +0.5\,\% with respect to that using the standard bin
width.

\item
The primary spectrum of the two electrons emitted in the \twonu\ decay of
\gess, which was then fed into the MC simulation, was sampled according to the
distribution given in Ref.~\cite{2nubbShape} implemented in
\textsc{Decay0}~\cite{decay0}.  The systematic uncertainty due to the assumed
\twonu\ spectral shape was evaluated by comparing the spectrum generated by
\textsc{Decay0} to the one given in Ref.~\cite{kotila}. Considering the
analysis window used for background modeling, the maximum deviation is 0.2\,\%
and the total deviation of the integral in the analysis window is 0.1\,\%.
When the fit with the background model is repeated using the spectrum of
Ref.~\cite{kotila}, the difference from the reference \thalftwo\ result is
less than 0.1\,\%.

\item
A possible effect of a transition layer, where it is assumed that the $n^{+}$
dead layer on the detector surfaces is partially active, has been
investigated~\cite{gerda-psd,trans-majo}. The dead layer thickness for
individual detectors assumed in MC simulations were given according to the
listed values in Ref.~\cite{background-paper}.  The transition layer is
modeled using two different assumptions: a linearly and an exponentially
increasing charge collection efficiency in the dead layer.  The systematic
uncertainty on \thalftwo\ due to the \twonu\ spectrum simulated with the
transition layer is found to be negligible.
\end{itemize}

\begin{figure*}[t]
\begin{center}
\includegraphics[width=0.73\textwidth]{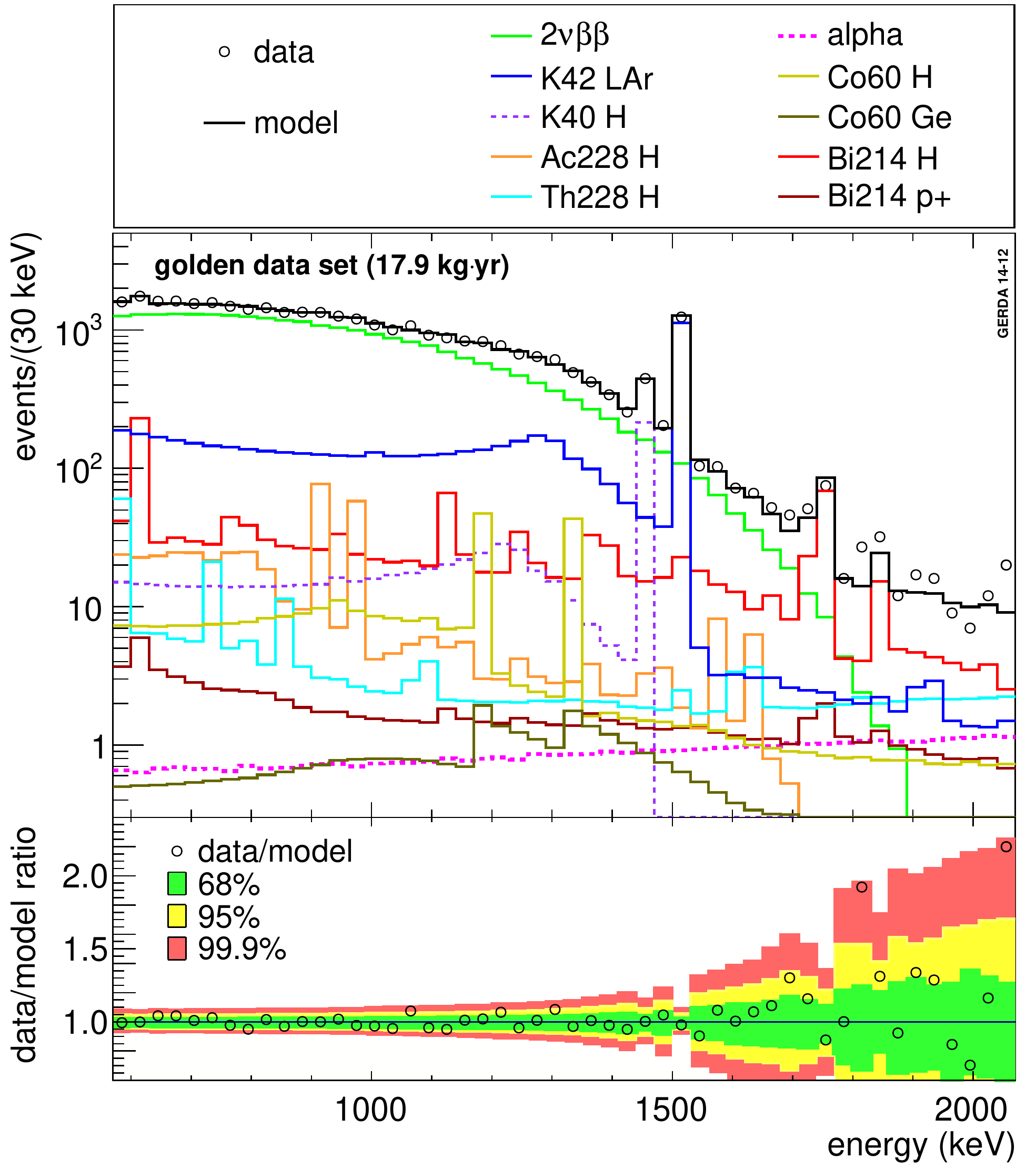}
\caption{\label{fig:twonu:result}
           Upper panel: experimental data (markers) and the best fit model
           (black histogram) for the golden data set. The contribution from
           \nnbb\ (green) and from the single background components are also
           shown.  Lower panel: ratio between experimental data and the
           prediction of the best fit model.  The green, yellow and red
           regions are the smallest intervals containing 68\,\%, 95\,\% and
           99\,\% probability for the ratio assuming the best fit parameters,
           respectively~\cite{Aggarwal2012}.
}
\end{center}
\end{figure*}

\noindent{\it (ii) MC simulation}

\noindent The uncertainty related to the MC simulation arises from the
precision of the experimental geometry model implemented in \mage\ (1\,\%) and
from the accuracy of particle tracking (2\,\%) performed by
\geant~\cite{geant4,geant-valid}.  The total MC simulation uncertainty was
estimated to be 2.2\,\% by summing in quadrature the aforementioned
contributions.\\

\noindent{\it (iii) Data acquisition and selection}

\noindent The trigger and reconstruction efficiencies for physical events are
practically 100\,\% above 100\,keV in \gerda. The performance of the quality
cuts applied in Phase~I data has been investigated through a visual
analysis. The total uncertainty related to data acquisition and selection was
estimated to be less than 0.1\,\%.

Summing in quadrature the uncertainties of the three groups gives a total
systematic uncertainty of $\pm$4.8\,\%.

\subsection{Results and Discussion}
\label{subsec:twonuResult}
Fig.~\ref{fig:twonu:result} shows the experimental data together with the best
fit model for the golden data set. The different components of the minimum
background model are also reported. The model is able to reproduce the
experimental data well, as shown in the lower panel of the figure by the
residuals.

The best estimate of the \thalftwo of the \twonu\ decay of \gess\ is:
\vspace*{5pt}
\begin{eqnarray}           \label{eq:twonu:result}
T^{2\nu}_{1/2} & = & \left ( 1.926\,^{+0.025}_{-0.022 \,\rm{stat}} \,^{+0.092}_{-0.092 \,\rm{syst}} \right) \cdot 10^{21}\,\rm{yr} 
\nonumber\\
&  = & (1.926 \pm 0.095) \cdot 10^{21}\,\rm{yr}\,,
\end{eqnarray} 
\vspace{5pt}

\noindent with the latter combining in quadrature the statistical (fit) and
systematic uncertainties.  The total uncertainty of 4.9\,\% is dominated by
the systematic uncertainties.  The largest contribution to the systematic
uncertainties comes from the uncertainty on the active \gess\ exposure
(4\,\%), which can only be reduced by performing new and more precise
measurements of the active masses of the coaxial detectors.  Other significant
contributions are related to the Monte Carlo simulations (2.2\,\%) and to the
background model assumptions ($^{+1.4\,\%} _{-1.2\,\%}$).  The latter have
been significantly reduced in this analysis compared to the analysis of the
first 5\,\kgyr\ of Phase~I data reported in Ref.~\cite{gerda-thalftwo}, where
the systematic uncertainty due to the background model was
$^{+5.7\,\%}_{-2.1\,\%}$.  The new result is in good agreement with that
mentioned above.  Adding further identified components to the reference
background model results in a slight increase of the best \thalftwo\ estimate.
 
The background level achieved in \gerda\ Phase~I is about one order of
magnitude lower with respect to predecessor \gess\ experiments, and has
allowed the measurement of \thalftwo\ with an unprecedented
signal-to-background ratio of 3:1 in the 570--2039~keV interval.  The ratio
amounts to 4:1 for the smaller interval of 600--1800~keV.
%

\section{Limits on Majoron-emitting double $\beta$-decays of $^{76}$Ge}
\label{sec:majoron}
\subsection{Analysis}
\label{subsec:majoronAnal}
The search for \onbbchi\ was performed using the golden and BEGe data sets,
amounting to a total exposure of 20.3~\kgyr. The analysis employed the
background model described in section~\ref{sec:background}. The information
from the two data sets was combined in one fit, while keeping their energy
spectra distinct. A separate fit was performed for each spectral index,
containing the background contributions, the contributions from \nnbb\ decay,
and also the Majoron component under study.  A single parameter, \thalfmajo,
is considered common for the two data sets.  It is defined as the half-life of
the respective Majoron accompanied mode.

In order to improve the detection efficiency for the Majoron processes with
low $n$ ($n = 1, 2$), a slightly different event selection was used with
respect to the \thalftwo\ analysis.  If an event occurs with energy deposition
in two detectors and the energy deposit in the detector where the decay took
place is below the threshold for the anti-coincidence cut, the event
contributes to the energy spectrum of the other detector. Therefore, when
determining the total energy spectrum resulting from decays in one of the
detectors, the energy spectra from all detectors in the array have to be taken
into account.  Such a selection has no impact on the detection efficiency for
the Majoron process with $n = 3 $ and 7 and \nnbb\ decay.  The content of the
$i$-th bin in the combined energy spectrum of all $N_{det}$ detectors in the
array, for decays taking place in the active and dead part of detector
$\alpha$, becomes:
\begin{eqnarray}\nonumber
\lambda^{\alpha, 0\nu\chi}_{i} & = & \frac { \left ( \rm{ln}\,2 \right )\,N_{A} }{ m_{\rm{enr}}\,
T^{0\nu\chi}_{1/2} } M_{\alpha} \,f_{76, \alpha} \cdot \left [ f_{AV, \alpha}\, \sum_{j = 1}^{N_{det}} \,t_{j} 
\varepsilon^{\alpha}_{AV,j}  \Phi^{\alpha,0\nu\chi}_{AV,i,j} \right. \\
&& \left. + (1-f_{AV, \alpha})\,\sum_{j = 1}^{N_{det}} \,t_{j} 
\varepsilon^{\alpha}_{DL,j}  \Phi^{\alpha,0\nu\chi}_{DL,i,j} \right ] \,
\label{eq:majoron:ThalfDerive}
\end{eqnarray}
with $\Phi^{\alpha,0\nu\chi}_{AV,i,j}$ ($\Phi^{\alpha,0\nu\chi}_{DL,i,j}$)
giving the content of the $i$-th bin of the normalized energy distribution
recorded with detector $j$ for \onbbchi\ taking place in the active (dead)
volume of detector $\alpha$.  Summing up the simulations of decays in all
$N_{det}$ detectors results in the final model spectrum:
\begin{equation}
\lambda^{0\nu\chi}_i = \sum^{N_{det}}_{\alpha=1} \lambda^{\alpha, 0\nu\chi}_{i} ~.
\end{equation}
For all four Majoron modes 
($n = 1, 2, 3, 7$) only lower limits on the half-life can be given. They were 
obtained from the 90\,\% quantiles of the marginalized posterior distributions. 
These lower limits for $T^{0\nu\chi}_{1/2}$, not taking into account the
systematic uncertainties, are in units of $10^{23}$ yr:
$>$4.4,      
$>$1.9,
$>$0.9, and
$>$0.4 for $n$ = 1, 2, 3, and 7, respectively.
The respective half-life of the \nnbb\ process derived from this analysis
amounts to  in units of $10^{21}$ yr: 
1.96$\pm0.03_{\rm{stat}}$, 
1.97$\pm0.03_{\rm{stat}}$,
1.98$\pm0.03_{\rm{stat}}$, and
1.99$\pm0.03_{\rm{stat}}$. 
Within the uncertainties coming from the different background models and the
different data sets of the two analyses, the derived \thalftwo\ values are in
agreement ($<$1\,$\sigma$) with that discussed in
section~\ref{subsec:twonuResult}.

\subsection{Systematic Uncertainties}
\label{subsec:majoronSyst}
The systematic uncertainties were divided into the three categories {\it (i)
  detector parameters and fit model}, {\it (ii) MC simulation}, and {\it (iii)
  data acquisition and selection}.

\vspace{5pt}
\noindent{\it (i) detector parameters and fit model}

\noindent   Uncertainties from the fitting procedure were folded
  into the posterior distribution of $T_{1/2}^{0\nu\chi}$ with a MC approach.
  Each source of uncertainty is described by a probability distribution.  The
  fitting procedure was repeated 1000 times, each time drawing a random number
  for each source of uncertainty according to its probability distribution:
\begin{itemize}
  \item
    Material screening measurement results were used to constrain the minimum
    number of events expected from close and medium distant sources of the
    $^{214}$Bi and $^{228}$Th decays.  Gaussian distributions describing these
    lower limits used in the fit were derived from the mean and standard
    deviations of the screening measurements.  For details refer to
    Ref.~\cite{background-paper}.
  \item
    As for the \thalftwo\ analysis, the standard fit uses a bin width of
    30~keV for the data and MC energy spectra. In order to determine the
    systematic uncertainty related to binning effects the bin width was
    sampled uniformly from 10 keV to 50 keV.
  \item
    Uncertainties on the active volume fractions enter the model in several
    ways.  On the one hand, the MC energy spectra for all internal sources,
    that is for \nnbb, \onbbchi, $^{60}$Co, and $^{68}$Ga decays, are
    affected, as the fraction of decays taking place in the active and dead
    part of the detectors changes with changing $f_{AV}$.  On the other hand,
    the uncertainty on the active volume fraction also plays a role for the
    shape of the energy spectrum due to $^{42}$K decays on the $n^+$
    surface. Larger $f_{AV}$ means thinner $n^+$ dead layer and thus the
    possibility of an increased contribution from the electrons to the
    spectrum.  For smaller $f_{AV}$ and thicker $n^+$ dead layer, their
    contributions are expected to be reduced.  The active volume fraction for
    each detector was sampled from a Gaussian distribution with mean and
    standard deviation according to Table~\ref{tab:detValues}.  For the
    coaxial detectors, the partial correlations of the uncertainty were taken
    into account.  The simulated spectra of the internal sources as well as of
    the $^{42}$K decays on the $n^+$ surface are composed according to the
    sampled active volume fractions.
  \item
    The uncertainty on the fraction of enrichment in $^{76}$Ge of the
    germanium that constitutes the detectors plays a role when converting the
    number of events attributed to \onbbchi\ into $T_{1/2}^{0\nu\chi}$.  The
    probability distribution of $f_{76}$ for each detector is given by a
    Gaussian function with mean values and standard deviations as listed in
    Table~\ref{tab:detValues}.
  \item
    The data does not allow the resolution of the ambiguity regarding the exact
    positions of the near and medium distant sources.  The $^{214}$Bi decays
    serves as a representative in order to estimate the impact of this
    uncertainty. Their near position is represented by decays in the holders,
    in the mini-shroud or on the $n^+$ surface of the detectors, each having a
    probability of $1/3$ in the sampling process.  The medium distant position
    is represented by decays in the radon-shroud or in the LAr, having a
    probability $1/2$ in contrast.
  \item
    Extensive studies of the characteristics of the BEGe diodes suggest the
    presence of a transition layer between the region where the detector is
    fully efficient and the external dead region~\cite{gerda-psd,trans-majo}.
    An uncertainty as high as $\pm$0.5\,\% on the lower limits of
    $T_{1/2}^{0\nu\chi}$ is estimated for this effect in the case of the BEGe
    detectors.  This uncertainty was folded into the total marginalized
    posterior distribution a posteriori.  The corresponding uncertainty for
    the coaxial detectors is estimated to be negligible.
\end{itemize}

The marginalized posterior distributions for $T_{1/2}^{0\nu\chi}$ derived from
each of the 1000 individual fits were sum\-med up.  The resulting total
marginalized posterior distribution accounts for the statistical as well as
for the listed systematic uncertainties related to the fit model.

As for the $T_{1/2}^{2\nu}$ analysis, the uncertainties on the active volume
fractions and on the enrichment fractions are major contributions to the total
uncertainty on the limits for $T_{1/2}^{0\nu\chi}$.  However, the largest
source of uncertainty is the composition of the fit model and the individual
background contributions.  In the case of $n=1$, a fit with a bin width of
50~keV weakens the limit by $\approx 16\,\%$ compared to the standard fit,
while the result for $T_{1/2}^{2\nu}$ is not affected at all. The stability of
the \thalftwo\ results shows the validity of the fit.  The use of the
alternative close and medium distant source positions for $^{214}$Bi decays
leads to maximal variations of $^{+8.3}_{-12.6}\,\%$ of the limit on
$T_{1/2}^{0\nu\chi}$.

\begin{table*} 
\begin{center}
\small 
\caption{  \label{tab:MajoronFinalResults}
          Experimental results for the limits on \thalfmajo\ of $^{76}$Ge for
          the Majoron models given in
          Refs.~\protect\cite{bam95,Hirsch1996,Carone1993,Mohapatra2000}.  The
          first section considers lepton number violating models (I) allowing
          $0\nu\beta\beta$ decay, while in the second section lepton number
          conserving models (II) are listed, where $0\nu\beta\beta$ decay is
          not allowed. The first column gives the model name, the second the
          spectral index, $n$, the third the information on whether one
          Majoron, $\chi$, or two Majorons, $\chi\chi$, is emitted, the fourth
          if the Majoron is a Goldstone boson, the fifth provides its lepton
          number, $L$, the sixth the experimental limit on
          $T_{1/2}^{0\nu\chi}$ of $^{76}$Ge obtained in this analysis.  The
          nuclear matrix elements, $\cal M$$^{0\nu\chi}$, the phase space
          factor, $G^{0\nu\chi}$, and the resulting effective coupling
          constants, $\langle g \rangle$, are given in the seventh, eighth and
          ninth columns, respectively.  The limits on $T_{1/2}^{0\nu\chi}$ of
          $^{76}$Ge for the Majoron models and $\langle g \rangle$ correspond
          to the $90\,\%$ quantiles of the marginalized posterior probability
          distribution.  For the case of $n=1$, the nuclear matrix element,
          $\cal M$$^{0\nu\chi}$, from
          Refs.~\protect\cite{Simkovic2013,Mustonen2013,Rodriguez2010,Menendez2009,Barea2013,Suhonen2010,Meroni2013}
          and the phase space factor, $G^{0\nu\chi}$, from
          Ref.~\protect\cite{Suhonen1998} are used for the calculation of
          $\langle g \rangle$.  The given range covers the variations of $\cal
          M$$^{0\nu\chi}$ in these works.  For $n=3\;\mathrm{and}\;7$,
          $\langle g \rangle$ is determined using the matrix elements and
          phase space factors from Ref.~\protect\cite{Hirsch1996}. The results
          for \onbbchi\ $(n=3,\;7)$ account for the uncertainty on ${\cal
            M}^{0\nu\chi}$.  For $n = 2$, only the experimental upper limit is
          given.
}
  \vspace{5pt}
  \begin{tabular*}{\textwidth}{@{\extracolsep{\fill}}l l l l l c c c c@{}}
    \hline
     Model  & n & Mode & Goldstone &$L$ & $T_{1/2}^{0\nu\chi}$ & $\cal M$$^{0\nu\chi}$ & $G^{0\nu\chi}$ & $\langle g \rangle$ \\
      &  &  & boson &  & [$10^{23}$\rm{yr}] &  & $[\mathrm{yr}^{-1}]$ &  \\
    \hline
    IB &  1 & $\chi$ & no & 0  & $>4.2$ & $(2.30-5.82)$ & $5.86 \cdot 10^{-17}$ & $< (3.4-8.7) \cdot 10^{-5}$\\
    IC &  1 & $\chi$ & yes & 0 & $>4.2$ & $(2.30-5.82)$ & $5.86 \cdot 10^{-17}$ & $< (3.4-8.7) \cdot 10^{-5}$\\
    ID &  3 & $\chi\chi$ & no & 0 &  $>0.8$ & $10^{-3 \pm 1}$ & $6.32 \cdot 10^{-19}$ & $< 2.1^{+4.5}_{-1.4}$\\
    IE &  3 & $\chi\chi$ & yes & 0 &  $>0.8$ & $10^{-3 \pm 1}$ & $6.32 \cdot 10^{-19}$ & $< 2.1^{+4.5}_{-1.4}$\\
    IF &  2 & $\chi$ & bulk field & 0 &  $>1.8$ & -- & -- & -- \\
    \hline
    IIB & 1 & $\chi$ & no & -2 & $>4.2$ & $(2.30-5.82)$ & $5.86 \cdot 10^{-17}$ & $< (3.4-8.7) \cdot 10^{-5}$\\ 
    IIC & 3 & $\chi$ & yes & -2 &  $>0.8$ & $0.16$ & $2.07 \cdot 10^{-19}$ & $< 4.7 \cdot 10^{-2}$ \\
    IID & 3 & $\chi\chi$ & no & -1 &  $>0.8$ & $10^{-3 \pm 1}$ & $6.32 \cdot 10^{-19}$ &$< 2.1^{+4.5}_{-1.4}$\\ 
    IIE & 7 & $\chi\chi$ & yes & -1 &  $>0.3$ & $10^{-3 \pm 1}$ & $1.21 \cdot 10^{-18}$ &$< 2.2^{+4.9}_{-1.4}$\\ 
    IIF & 3 & $\chi$ & gauge boson & -2  & $>0.8$ & $0.16$ & $2.07 \cdot 10^{-19}$ & $< 4.7 \cdot 10^{-2}$\\
    \hline
  \end{tabular*}
\end{center}
\end{table*}

\vspace{5pt}

\noindent{\it (ii) MC simulation}

\noindent As in the case of the $T_{1/2}^{2\nu}$ measurement, a total MC
simulation uncertainty of 2.2\,\% has to be taken into account for effects
related to the geometry implementation and particle tracking.  It is folded
into the total marginalized posterior distributions.  No effect on the lower
limits is observed for any of the spectral modes.

\vspace{5pt}

\noindent{\it (iii) Data acquisition and selection}

\noindent The uncertainty from data acquisition and selection is estimated to
be below $0.1\,\%$ and does not alter the derived limits on
$T_{1/2}^{0\nu\chi}$.

\subsection{Results and Discussion}
\label{subsec:majoronResult}
\begin{figure*}
\centering
\includegraphics[width=0.75\textwidth]{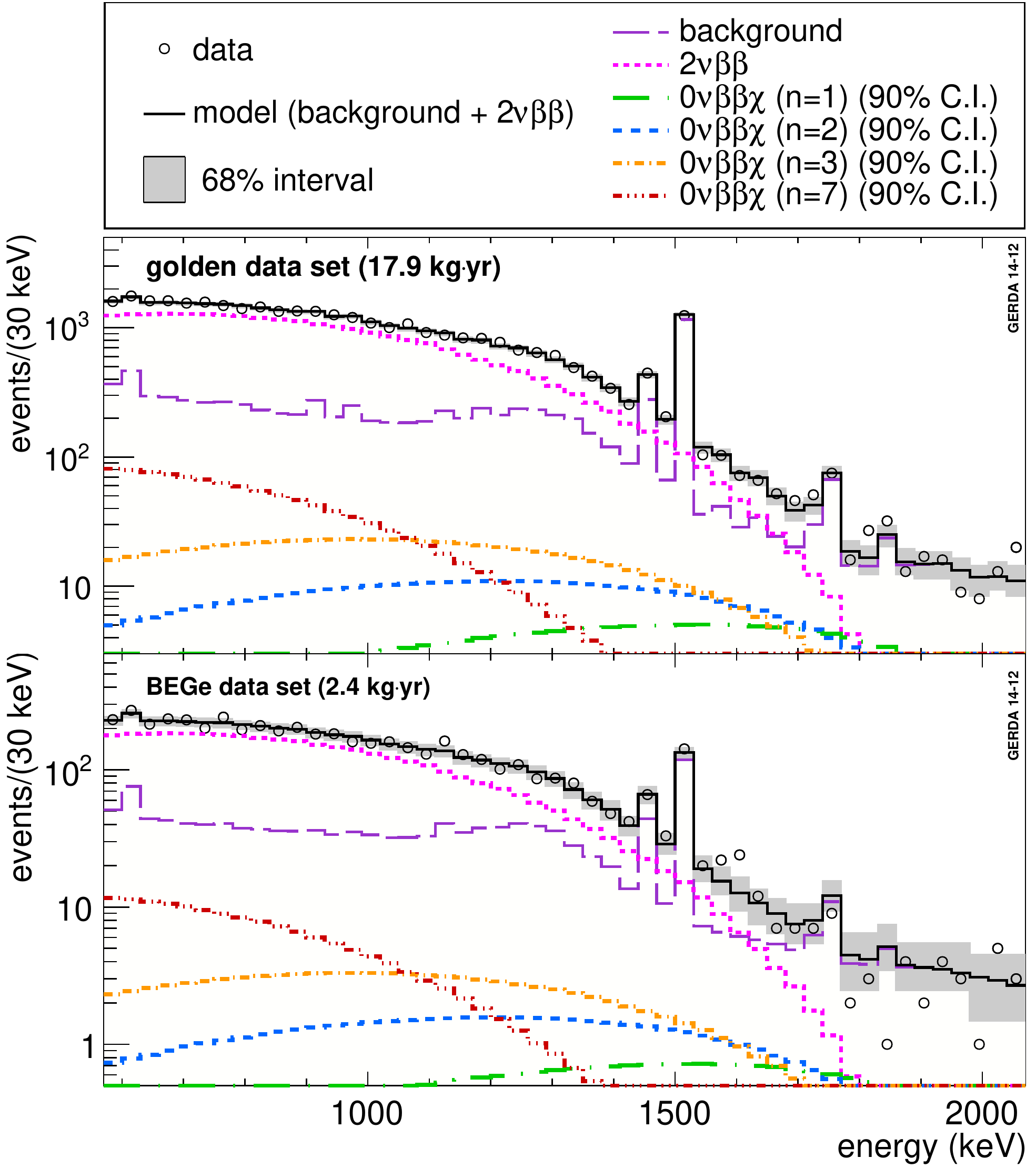}
\caption{\label{fig:thalfmajo}
       Best-fit model and data energy spectrum for the coaxial and the BEGe
       data sets for the case of spectral index $n = 1$. The contributions
       from \nnbb\ decay and the background contributions are shown
       separately. The best-fit model does not contain the contributions from
       \onbbchi. The smallest interval of 68\,\% probability for
       the model expectation is indicated in grey. Also shown is the upper
       limit for \onbbchi\ decay with $n = 1$ as determined from
       the 90\,\% quantile of the marginalized posterior probability for
       1/\thalfmajo. For illustrative purpose, also the upper limits at 90\,\%
       C.I. of the other three spectral indices $n = 2, 3, 7$ are reported.
}
\end{figure*}

Fig.~\ref{fig:thalfmajo} shows the global model for the case of spectral index
$n = 1$ together with the energy spectra for both the coaxial and the BEGe
data sets.  The contributions from the background contaminations, from the
\nnbb\ decay only, and the combined spectra from the background contaminations
and \nnbb\ decay are drawn separately.  The 35868 events in the data spectrum
of the golden data set were matched with 35834 events in the best-fit model
for $n = 1$. Of those events, in the best fit, 54.5 are attributed to
\onbbchi. For the BEGe data set, the best-fit model contains 5081.4 counts for
the 5035 measured events. In this fit, 7.8 events are attributed to
\onbbchi\ decay.  The limit of \thalfmajo\ at 90\,\%~C.I. derived from the fit
is also drawn (green histogram). The upper limits at 90\,\%~C.I. for the
remaining three modes are reported for illustrative purpose (blue histogram
for $n = 2$, orange for $n = 3$ and red for $n = 7$). The maximum of the
corresponding distributions shifts to higher energy with the diminishing of
the spectral index $n$.  The resulting lower limits on \thalfmajo, determined
as the 90\,\% quantiles of the posterior probability distributions and taking
into account all uncertainties related to the fit model, are (in units of
10$^{23}$ yr): $>$4.2, $>$1.8, $>$0.8 and $>$0.3 for $n=1,2,3$ and 7,
respectively.  The results are summarized in
Table~\ref{tab:MajoronFinalResults} for the different Majoron models.

The limits on \thalfmajo\ presented here are the most stringent limits
obtained to date for $^{76}$Ge. The limits for $n = 1$ and $n = 3 $ are
improved by more than a factor six~\cite{HdM-latest}, the limit for $n = 7$ is
improved by a factor five~\cite{HdM} compared to previous measurements. The
limit for the mode with $n = 2$ is reported here for the first time.
 
From the lower limits on \thalfmajo, upper limits on the effective
neutrino-Majoron coupling constants $\langle g \rangle$ for the models with $n
= 1, 3$ and 7 can be calculated using the following equations:
\begin{equation}
1/T^{0\nu\chi}_{1/2} = |\langle g \rangle|^2 \cdot 
G^{0\nu\chi} (Q_{\beta\beta}, Z) \cdot |M^{0 \nu \chi}|^2
\end{equation}
and 
\begin{equation}
1/T^{0\nu\chi}_{1/2} = |\langle g \rangle|^4 \cdot 
G^{0\nu\chi\chi} (Q_{\beta\beta}, Z) \cdot |M^{0\nu\chi\chi}|^2 
\end{equation}
for single and double Majoron emission, respectively.  The matrix element for
the models with $n =1$ (IB, IC and IIB) are taken from
Refs.~\cite{Simkovic2013,Mustonen2013,Rodriguez2010,Menendez2009,Barea2013,Suhonen2010,Meroni2013},
whereas the phase space factor is that of Ref.~\cite{Suhonen1998}.  The matrix
elements for the models with $n = 3$ (ID, IE, IIC, IID, IIF) and with $n=7$
(IIE) as well as the corresponding phase space factors are taken from
Ref.~\cite{Hirsch1996}.  The results for the upper limits on $\langle g
\rangle$ are also shown in Table~\ref{tab:MajoronFinalResults}.  The coupling
constants allow a comparison with other isotopes.  The best limits on
\onbbchi\ decay of isotopes other than $^{76}$Ge have been obtained for
$^{100}$Mo~\cite{NEMO2006} and $^{136}$Xe~\cite{KamLAND-Zen}.  When comparing
with the case of $^{100}$Mo, it becomes obvious that the limits on
\thalfmajo\ determined in the present analysis are about one order of
magnitude more stringent, for the case of $n = 7$ even two orders of
magnitude. However, due to the differences in the matrix elements and the
phase space factors, the resulting limits on $\langle g \rangle$ from
$^{100}$Mo and $^{76}$Ge are comparable. The limits for $\langle g \rangle$
derived from $^{136}$Xe are a factor of two to five more stringent due to the
higher limits that had been measured for \thalfmajo.

\section{Conclusions}
\label{sec:conclusions}
Phase~I of the \gerda\ experiment, located at the INFN Laboratori Nazionali
del Gran Sasso (LNGS) in Italy, has been executed between November 2011 and
May 2013. Utilizing the collected exposure of Phase~I, an improved result of
the half-life of the \nnbb\ process in $^{76}$Ge was obtained and new limits
for the half-lives of the Majoron-emitting double beta decays were produced.

The half-life for the \nnbb\ process is determined to be:
\begin{equation}
T^{2 \nu}_{1/2} = (1.926 \pm 0.095) \cdot 10^{21} ~\textrm{yr}\,\,.
\end{equation}
Thanks to the extremely low background level in the \gerda\ experiment, with a
signal-to-background ratio of 3:1 in the 570--2039~keV interval and a refined
background model, the measurement has an unprecedented precision ($<$5\,\%)
with respect to previous experiments using $^{76}$Ge.  The new result is in
good agreement with the one derived from a smaller data set with
5~\kgyr\ exposure~\cite{gerda-thalftwo}.  The inclusion of more components
into the reference background model results in a slight increase of the best
estimate for \thalftwo.

Majoron emission processes were searched for in the energy spectra using an
exposure of 20.3~kg$\cdot$yr.  The analysis was performed for all four
possibilities of the spectral index $n$ ($ n = $ 1, 2, 3, and 7).  No
indication for a contribution of $0\nu \beta\beta \chi$ was found in any of
the cases.  Lower limits on the half-lives, \thalfmajo, were determined from
the quantiles of 90\,\% probability of the marginalized posterior probability
distributions. The results constitute the most stringent limits on
\thalfmajo\ of $^{76}$Ge obtained to date.  For the standard mode ($n = 1$),
the lower limit is determined to be:
\begin{equation}
T^{0 \nu \chi}_{1/2} > 4.2 \cdot 10^{23} ~\textrm{yr}.
\end{equation}
From the lower limit on \thalfmajo, an upper limit on the effective 
neutrino-Majoron coupling constant, $\langle g \rangle$, can be 
inferred:
\begin{equation}
\langle g \rangle < (3.4 - 8.7) \cdot 10^{-5}.
\end{equation}

\section*{Acknowledgments}
 The \gerda\ experiment is supported financially by
   the German Federal Ministry for Education and Research (BMBF),
   the German Research Foundation (DFG) via the Excellence Cluster Universe,
   the Italian Istituto Nazionale di Fisica Nucleare (INFN),
   the Max Planck Society (MPG),
   the Polish National Science Centre\\  (NCN),
   the Foundation for Polish Science (MPD programme),
   the Russian Foundation for Basic Research (RFBR), and
   the Swiss National Science Foundation (SNF).
 The institutions acknowledge also internal financial support.

The \gerda\ Collaboration thanks the directors and the staff of the LNGS
for their continuous strong support of the \gerda\ experiment.


\end{document}